\documentclass[12pt]{article}

\def\case#1#2{{\textstyle{#1\over #2}}}
\def\sech{\mathop{\rm sech}\nolimits}
\def\cosech{\mathop{\rm cosech}\nolimits}

\begin{document}

\title{Non-Hermitian Hamiltonians with real and complex eigenvalues: An sl(2,C)
approach}
\author{B. Bagchi $^{a}$ and
C. Quesne $^{b}$\\ 
$^a$ {\small Department of Applied Mathematics, University of Calcutta, India}\\ 
$^b$ {\small Physique Nucl\'eaire Th\'eorique et Physique
Math\'ematique,  Universit\'e Libre de Bruxelles,}\\
{\small Belgium}}
\date{ }
%
%
%
\maketitle

\begin{abstract} Potential algebras are extended from Hermitian to non-Hermitian
Hamiltonians and shown to provide an elegant method for studying the transition from
real to complex eigenvalues for a class of non-Hermitian Hamiltonians associated with
the complex Lie algebra A$_1$.
\end{abstract}
%
%
\section{Introduction}

In recent years there has been much interest in non-Hermitian Hamiltonians with real,
bound-state eigenvalues. In particular, PT-symmetric Hamiltonians (such that $(PT) H
(PT)^{-1} = H$, where $P$ is the parity and $T$ the time reversal) have been
conjectured to have a real bound-state spectrum except when the symmetry is
spontaneously broken, in which case their complex eigenvalues should come in conjugate
pairs~\cite{bender}. It is also known that PT symmetry is not a necessary condition for
the occurrence of real or complex-conjugate pairs of eigenvalues. A more general
condition, namely pseudo-Hermiticity of the Hamiltonian (i.e., the existence of a Hermitian
linear automorphism $\eta$ such that
$\eta H \eta^{\dagger} = H^{\dagger}$) has been identified as an explanation for the
existence of this phenomenon for some non-PT-symmetric Hamiltonians~\cite{mosta,
ahmed01b}. It should be noted that all such non-Hermitian Hamiltonians require a
generalization of the normalization condition corresponding to an indefinite scalar product
(see~\cite{bagchi01a} and references quoted therein).\par
%
%
Very recently there has been a growing interest in determining the critical strengths of
the interaction, if any, at which PT symmetry (or some generalization thereof) becomes
spontaneously broken. Among the various techniques that have been employed to
contruct and study non-Hermitian Hamiltonians with real or complex spectra, algebraic
methods provide powerful approaches. In the present communication, we show how
potential algebras can be used for such a purpose.\par
%
%
\section{Potential algebras for Hermitian and non-Hermitian Hamiltonians}

Potential algebras refer to Lie algebras whose generators connect eigenfunctions
$\psi^{(m)}_n(x)$ corresponding to the same eigenvalue (i.e., $E^{(m)}_n$ is
constant), but to different potentials $V_m(x)$ of a given family. Here $m$ is some
parameter (generally related to the potential strength), which may change by one unit
under the action of the generators.\par
%
%
Potential algebras were introduced for Hermitian Hamiltonians as real Lie algebras, the
simplest example being that of $\mbox{\rm sl(2,R)} \simeq
\mbox{so(2, 1)}$~\cite{alhassid, frank}. The latter is generated by $J_0$, $J_+$, $J_-$,
satisfying the commutation relations
\begin{equation}
  [J_0, J_{\pm}] = \pm J_{\pm}, \qquad [J_+, J_-] = - 2J_0, \label{eq:commutation}  
\end{equation}
and the Hermiticity properties $J_0^{\dagger} = J_0$, $J_{\pm}^{\dagger} =
J_{\mp}$. Such operators are realized as differential operators
\begin{equation}
  J_0 = - {\rm i} \frac{\partial}{\partial\phi}, \qquad J_{\pm} = e^{\pm{\rm i}\phi}
  \left[\pm\frac{\partial}{\partial x} + \left({\rm i} \frac{\partial}{\partial\phi}
  \mp \frac{1}{2}\right) F(x) + G(x)\right],  \label{eq:J}  
\end{equation}
depending upon a real variable $x$ and an auxiliary variable $\phi \in [0, 2\pi)$,
provided the two real-valued functions $F(x)$ and $G(x)$ in (\ref{eq:J}) satisfy
coupled differential equations
\begin{equation}
  F' = 1 - F^2, \qquad G' = - FG.  \label{eq:F-G-diff}  
\end{equation}
The sl(2,R) Casimir operator, $J^2 = J_0^2 \mp J_0 - J_{\pm} J_{\mp}$, then
becomes a second-order differential operator.\par
%
%
{}For bound states, to which we restrict ourselves here, one considers unitary irreducible
representations of sl(2,R) of the type $D^+_k$, spanned by states $|km\rangle$, $k \in
\mbox{\rm R}^+$, $m = k+n$, $n \in \mbox{\rm N}$, such that $J_0 |km\rangle = m
|km\rangle$ and $J^2 |km\rangle = k(k-1) |km\rangle$. In the realization (\ref{eq:J}),
these states are given by $|km\rangle = \Psi_{km}(x, \phi) = \psi_{km}(x) e^{{\rm i}
m\phi}/\sqrt{2\pi}$, where $\psi_{km}(x) = \psi^{(m)}_n(x)$ satisfies the
Schr\"odinger equation
\begin{equation}
  - \psi^{(m)\prime\prime}_n + V_m \psi^{(m)}_n = E^{(m)}_n \psi^{(m)}_n.
  \label{eq:SE}
\end{equation}
In (\ref{eq:SE}), the potential $V_m(x)$ is defined in terms of $F(x)$ and $G(x)$ by 
\begin{equation}
  V_m = \left(\case{1}{4} - m^2\right) F' + 2m G' + G^2,  \label{eq:V}  
\end{equation}
and the energy eigenvalues are given by $E^{(m)}_n = - \left(m - n -
\case{1}{2}\right)^2$. The eigenfunction $\psi^{(m)}_0(x)$ can be easily constructed
by solving the first-order differential equation $J_- \psi^{(m)}_0(x) = 0$, while the
remaining eigenfunctions $\psi^{(m)}_n(x)$ can be obtained from the action of $J_+$
on $\psi^{(m-1)}_{n-1}(x)$. Imposing the regularity condition for bound states
$\psi^{(m)}_n(\pm\infty) \to 0$ restricts the allowed values of $n$ to $n = 0, 1,
\ldots, n_{\rm max} < m - \frac{1}{2}$.\par
%
%
Detailed inspection of the system of differential equations (\ref{eq:F-G-diff}) shows that
it admits three classes of solutions corresponding to the (nonsingular) Scarf II, the
(singular) generalized P\"oschl-Teller, and the Morse potentials,
respectively~\cite{wu,englefield}.\par
%
%
The transition from Hermitian to non-Hermitian Hamiltonians can now be performed by
replacing real Lie algebras by complex ones~\cite{bagchi00, bagchi02} (see
also~\cite{levai00, levai01} for a related approach). In the case of sl(2,R), we find the
algebra known as A$_1$ in Cartan's classification of complex Lie algebras (which we shall
refer to as sl(2,C), considered here as a complex algebra, not a real one as is usely the
case). Its generators still satisfy the commutation relations (\ref{eq:commutation}), but
their Hermiticity properties remain undefined. This means that the realization (\ref{eq:J})
is still applicable with $F(x)$ and $G(x)$ now some complex-valued functions.\par
%
%
{}For bound states, we restrict ourselves to irreducible representations spanned by
states $|km\rangle$, for which both $k = k_R + {\rm i} k_I$ and $m = m_R + {\rm i} m_I$
may be complex with $m_R = k_R + n \in \mbox{\rm R}$, $m_I = k_I \in \mbox{\rm
R}$, and $n \in \mbox{\rm N}$. Equations (\ref{eq:SE}) and (\ref{eq:V}) remain valid
and we get some complexified forms of the Scarf II, generalized P\"oschl-Teller, and Morse
potentials, given by 
\begin{eqnarray}
  {\rm I:\quad} & V_m & =  \left(b^2 - m^2+ \case{1}{4}\right) \sech^2\tau - 2 mb
         \sech\tau \tanh\tau, \label{eq:VI}  \\[0.1cm]
  {\rm II:\quad} & V_m & =  \left(b^2 + m^2 - \case{1}{4}\right) \cosech^2\tau
         - 2mb \cosech\tau \coth\tau, \label{eq:VII} \\[0.1cm]         
  {\rm III:\quad} & V_m & =  b^2 e^{\mp 2x} \mp 2mb e^{\mp x}, 
         \label{eq:VIII}
\end{eqnarray}
where $b = b_R + {\rm i}b_I$ and $\tau = x-c-{\rm i}\gamma$ with $b_R$, $b_I$, $c
\in \mbox{\rm R}$, $-\frac{\pi}{4} \le \gamma < \frac{\pi}{4}$. For generic values of
the parameters, such potentials are neither PT-symmetric nor pseudo-Hermitian. The
corresponding energy eigenvalues become $E^{(m)}_n = - \left(m_R + {\rm i}m_I - n -
\case{1}{2}\right)^2$ and are therefore complex if $m_I \ne 0$. From the explicit
form of $\psi^{(m)}_0(x)$, it can be shown that the potentials (\ref{eq:VI}) --
(\ref{eq:VIII}) have at least one regular eigenfunction (namely that corresponding to
$n=0$) provided $m_R > 1/2$ and $b_R > 0$, where the second condition applies only
to class III.\par
%
%
It is worth noting that the same potentials (\ref{eq:VI}) -- (\ref{eq:VIII}) can alternatively
be derived in the framework of supersymmetric quantum mechanics by considering a
complex superpotential $W(x) = \left(m - \frac{1}{2}\right) F(x)
- G(x)$~\cite{bagchi01b}. They turn out to be shape-invariant as their real counterparts.
%
%
\section{Some examples}

As a first example, let us consider a special case of class I potentials,
\begin{equation}
  V(x) = - V_1 \sech^2 x - {\rm i} V_2 \sech x \tanh x, \qquad V_1 > 0, \qquad
  V_2 \ne 0,  \label{eq:Scarf}
\end{equation}
which is both PT-symmetric and P-pseudo-Hermitian. It corresponds to $c = \gamma =
0$ in (\ref{eq:VI}), while the other four parameters $m_R$, $m_I$, $b_R$,
$b_I$ are related to $V_1$ and $V_2$ through four quadratic equations,
\begin{eqnarray}
  && b_R^2 - b_I^2 - m_R^2 + m_I^2 + \case{1}{4} = - V_1,  \\
  && b_R b_I - m_R m_I= 0, \\
  && m_R b_R - m_I b_I = 0,  \\
  && 2(m_R b_I + m_I b_R)  = V_2. 
\end{eqnarray}
\par
%
%
On solving the latter to express $m_R$, $m_I$, $b_R$, $b_I$ in terms of $V_1$,
$V_2$, taking the regularity condition $m_R > 1/2$ into account, and inserting the
$m_R$ and $m_I$ values into $E^{(m)}_n$, we find one critical strength (for a given sign
of $V_2$) corresponding to $|V_2| = V_1 + \frac{1}{4}$. For $|V_2| < V_1 +
\frac{1}{4}$, there are (in general) two series of real eigenvalues (instead of one for the
real Scarf II potential), given by
\begin{equation}
  E_{n,\pm} = - \left[\case{1}{2} \left(\sqrt{V_1 + \case{1}{4} + |V_2|} \pm 
  \sqrt{V_1 + \case{1}{4} - |V_2|}\right) - n - \case{1}{2}\right]^2, 
\end{equation}
where $n= 0$, 1, 2,~\ldots ${}< \frac{1}{2} \left(\sqrt{V_1 + 
\frac{1}{4} + |V_2|} \pm \sqrt{V_1 + \frac{1}{4} - |V_2|} - 1\right)$. When
$|V_2|$ reaches the value $V_1 + \frac{1}{4}$, the two series of real energy
eigenvalues merge and for higher $|V_2|$ values they move into the complex plane so
that we get a series of complex-conjugate pairs of eigenvalues,
\begin{equation}
  E_{n,\pm} = - \left[\case{1}{2} \left(\sqrt{|V_2| + V_1 + \case{1}{4}} \pm {\rm i}
  \sqrt{|V_2| - V_1 - \case{1}{4}}\right) - n - \case{1}{2}\right]^2, 
\end{equation}
where $n = 0$, 1, 2,~\ldots ${}< \case{1}{2} \left(\sqrt{|V_2| + V_1 + \case{1}{4}} -
1\right)$. This agrees with results obtained elsewhere by another
method~\cite{ahmed01a}.\par
%
%
Another example of occurrence of just one critical strength is provided by a
PT-symmetric and P-pseudo-Hermitian special case of class II potentials,
\begin{equation}
  V(x) = V_1 \cosech^2\tau - V_2 \cosech\tau \coth\tau, \qquad
  V_1 > - \case{1}{4}, \qquad V_2 \ne 0,  
\end{equation}
which is defined on the entire real line (contrary to its real counterpart).\par
%
%
A rather different situation is depicted by the potential
\begin{equation}
  V(x) = (V_{1R} + {\rm i}V_{1I}) e^{-2x} - (V_{2R} + {\rm i}V_{2I}) e^{-x}, \qquad
  V_{1R}, V_{1I}, V_{2R}, V_{2I} \in \mbox{\rm R},  \label{eq:Morse}
\end{equation}
which is the most general class III potential (for the upper sign choice in (\ref{eq:VIII}))
and has no special property for generic values of the parameters.\par
%
%
By proceeding as in the first example, we find that the regularity conditions $m_R >
1/2$ and $b_R >0$ impose that $V_{1I}$ be non-vanishing and 
\begin{equation}
  (V_{1R} + \Delta)^{1/2} V_{2R} + \nu (- V_{1R} + \Delta)^{1/2} V_{2I} > \sqrt{2}
  \Delta,  \label{eq:Morse-cond}
\end{equation}
where $\nu$ denotes the sign of $V_{1I}$ and $\Delta \equiv \sqrt{V_{1R}^2 +
V_{1I}^2}$.\par
%
%
The results for the bound-state energy eigenvalues strongly contrast with those
obtained hereabove. Indeed real eigenvalues belonging to a single series,
\begin{equation}
  E_n = - \left[\frac{V_{2R}}{\sqrt{2}|V_{1I}|} (- V_{1R} + \Delta)^{1/2} - n -
  \frac{1}{2}\right]^2, 
\end{equation}
where $n=0$, 1, 2,~\ldots ${}< (V_{2R}/\sqrt{2}|V_{1I}|) (- V_{1R} + \Delta)^{1/2} -
\frac{1}{2}$, only occur for a special value of $V_{2I}$, namely $V_{2I} = \nu (-
V_{1R} + \Delta)^{1/2} (V_{1R} + \Delta)^{-1/2} V_{2R}$,  
while complex eigenvalues,
\begin{eqnarray}
  E_n & = & - \Biggl\{\frac{1}{2\sqrt{2} \Delta} \left[(V_{1R} + \Delta)^{1/2} - {\rm i}
\nu
  (- V_{1R} + \Delta)^{1/2}\right] (V_{2R} + {\rm i}V_{2I}) \nonumber \\
  && \mbox{} - n - \frac{1}{2}\Biggr\}^2,
\end{eqnarray}
where $n = 0$, 1, 2,~\ldots ${}< \frac{1}{2\sqrt{2} \Delta} \left[(V_{1R} + 
\Delta)^{1/2} V_{2R} + \nu (- V_{1R} + \Delta)^{1/2} V_{2I}\right] -
\frac{1}{2}$ and which do not form complex-conjugate pairs, occur for all the
remaining values of $V_{2I}$.\par
%
%
Such results can be interpreted by choosing the parametrization $V_{1R} = A^2 -
B^2$, $V_{1I} = 2AB$, $V_{2R} = \gamma A$, $V_{2I} = \delta B$, where $A$, $B$,
$\gamma$, $\delta$ are real, $A>0$, and $B \ne 0$. The complexified Morse potential
(\ref{eq:Morse}) then becomes $V(x) = (A + {\rm i}B)^2 e^{-2x} - (2C+1) (A + {\rm i}B)
e^{-x}$,  where $C = [(\gamma-1) A + {\rm i} (\delta-1) B]/[2(A + {\rm i} B)]$. Its (real
or complex) eigenvalues can be written in a unified way as $E_n = - (C-n)^2$, while the
regularity condition (\ref{eq:Morse-cond}) amounts to
$(\gamma-1) A^2 + (\delta-1) B^2 > 0$.\par
%
%
{}For $\delta = \gamma > 1$, and therefore $C = \frac{1}{2} (\gamma-1) \in
\mbox{\rm R}^+$, $V(x)$ is pseudo-Hermitian under imaginary shift of the
coordinate~\cite{ahmed01b}. It has only real eigenvalues corresponding to
$n=0$, 1, 2,~\ldots ${}< C$, thus exhibiting no symmetry breaking over the whole
parameter range. For the values of $\delta$ different from $\gamma$, the potential
indeed fails to be pseudo-Hermitian. In such a case, $C$ is complex as well as the
eigenvalues. Nevertheless, the eigenfunctions associated with $n=0$, 1, 2,~\ldots ${}<
{\rm Re}\,C$ remain regular. Note that the existence of regular eigenfunctions with
complex energies for general complex potentials is a well-known phenomenon (see
e.g.~\cite{baye}).\par
%
%

\end{document}